\PassOptionsToPackage{unicode}{hyperref}
\PassOptionsToPackage{hyphens}{url}
\documentclass[
  11pt,
]{article}
\usepackage{amsmath,amssymb}
\usepackage{iftex}
\ifPDFTeX
  \usepackage[T1]{fontenc}
  \usepackage[utf8]{inputenc}
  \usepackage{textcomp} % provide euro and other symbols
\else % if luatex or xetex
  \usepackage{unicode-math} % this also loads fontspec
  \defaultfontfeatures{Scale=MatchLowercase}
  \defaultfontfeatures[\rmfamily]{Ligatures=TeX,Scale=1}
\fi
\usepackage{lmodern}
\ifPDFTeX\else
\fi
\IfFileExists{upquote.sty}{\usepackage{upquote}}{}
\IfFileExists{microtype.sty}{% use microtype if available
  \usepackage[]{microtype}
  \UseMicrotypeSet[protrusion]{basicmath} % disable protrusion for tt fonts
}{}
\usepackage{xcolor}
\usepackage{longtable,booktabs,array}
\usepackage{calc} % for calculating minipage widths
\usepackage{etoolbox}
\makeatletter
\patchcmd\longtable{\par}{\if@noskipsec\mbox{}\fi\par}{}{}
\makeatother
\IfFileExists{footnotehyper.sty}{\usepackage{footnotehyper}}{\usepackage{footnote}}
\makesavenoteenv{longtable}
\usepackage{graphicx}
\makeatletter
\def\maxwidth{\ifdim\Gin@nat@width>\linewidth\linewidth\else\Gin@nat@width\fi}
\def\maxheight{\ifdim\Gin@nat@height>\textheight\textheight\else\Gin@nat@height\fi}
\makeatother
\setkeys{Gin}{width=\maxwidth,height=\maxheight,keepaspectratio}
\makeatletter
\def\fps@figure{htbp}
\makeatother
\providecommand{\tightlist}{%
  \setlength{\itemsep}{0pt}\setlength{\parskip}{0pt}}
\usepackage{pos}

\author[a]{Christopher Aubin}
\affiliation[a]{Department of Physics \& Engineering Physics, Fordham University, Bronx, New York, NY 10458, USA}

\author[b]{Bipasha Chakraborty}
\affiliation[b]{University of Southampton, SO17 IBJ, UK}

\author[c,d]{Will~Detmold}
\affiliation[c]{Center for Theoretical Physics, Massachusetts Institute of Technology, Cambridge, MA 02139, USA}
\affiliation[d]{The NSF Institute for Artificial Intelligence and Fundamental Interactions}

\author[e,f]{Sofie Martins}
\affiliation[e]{IMADA, University of Southern Denmark, Campusvej 55, 5230 Odense M, Denmark}
\affiliation[f]{$\hbar$QTC, Quantum Theory Center, University of Southern Denmark, Campusvej 55, 5230 Odense M, Denmark}

\author[g]{Nilmani Mathur}
\affiliation[g]{Department of Theoretical Physics, Tata Institute of Fundamental Research, Colaba, Mumbai 400005, India}

\author[h]{Tereza Mendes}
\affiliation[h]{Instituto de Física de São Carlos, Universidade de São Paulo}

\author[i]{Finn~M.~Stokes}
\affiliation[i]{J{\"u}lich Supercomputing Centre, Forschungszentrum J{\"u}lich, J{\"u}lich D-52428, Germany}

\onbehalf{on behalf of the Lattice Diversity and Inclusion Committee}

\emailAdd{lattice\_diversity@mit.edu}
\FullConference{
The 40th International Symposium on Lattice Field Theory (Lattice 2023)\\
	July 31st - August 4th, 2023\\
	Fermi National Accelerator Laboratory
}

\abstract{We review the level of welcomeness that members of the lattice field theory community feel based on the results of a survey performed in May and June 2023. While respondents reported generally high levels of feeling welcome at the lattice conference, women and people with diverse gender identities, sexual orientations, ethnic backgrounds and religious affiliations feel less included and have more negative experiences at the lattice conference than their peers. Respondents report that they are actively informing themselves about inequities in the community, however a large fraction of survey participants underestimate the severity of the problem, as was found in previous surveys. The survey data indicate that this situation can be most effectively improved by organizing talks and events about issues of diversity and inclusion within the lattice community. Respondents also reported that individual readings of scientific papers on equity and inclusion are effective in giving people agency in making a change and hence it may be helpful to collate a collection of important articles on these topics.}

\usepackage{flafter}
\usepackage{pos}
\usepackage{natbib}
\usepackage{booktabs}
\usepackage{longtable}
\usepackage{array}
\usepackage{multirow}
\usepackage{wrapfig}
\usepackage{float}
\usepackage{colortbl}
\usepackage{pdflscape}
\usepackage{tabu}
\usepackage{threeparttable}
\usepackage{threeparttablex}
\usepackage[normalem]{ulem}
\usepackage{makecell}
\usepackage{xcolor}
\ifLuaTeX
  \usepackage{selnolig}  % disable illegal ligatures
\fi
\usepackage[]{natbib}
\nocite{*}
\IfFileExists{bookmark.sty}{\usepackage{bookmark}}{\usepackage{hyperref}}
\IfFileExists{xurl.sty}{\usepackage{xurl}}{} % add URL line breaks if available
\hypersetup{
  pdftitle={LDIC Survey 2023: Feeling Welcome in the Community},
  hidelinks,
  pdfcreator={LaTeX via pandoc}}

\title{LDIC Survey 2023: Feeling Welcome in the Community}
\date{\vspace{-2.5em}2023-07-19}

\begin{document}
\maketitle

{
\setcounter{tocdepth}{2}
\tableofcontents
}
Find out more about the activities of the Lattice Diversity and Inclusion Committee at

\[\text{\texttt{latticediversity.github.io}}\]

or write us at \texttt{lattice\_diversity@mit.edu}.

\hypertarget{motivation}{%
\section{Motivation}\label{motivation}}

The LDIC has distributed regular surveys to understand whether every member of our community feels equally welcome at conferences, in particular at the lattice conference, both during scientific discussions and other portions of the event. For this survey, we inquired how the respondents perceived the social climate at past conferences. We asked about their subjective experience, any negative incidents that occurred and whether they felt comfortable to discuss them. We additionally asked more general questions about how equitable they perceive the lattice field theory community to be and whether they feel that there is sufficient information available to allow them to correct existing inequities, if they are able to.

Previous studies of diversity in the lattice field theory community can be found in \citep{Aubin:2019rdf}, \citep{Lin:2016age} and \citep{poster:2021}.

\hypertarget{study-design}{%
\section{Study Design}\label{study-design}}

The survey was conducted using Google Forms, advertised through two e-mails over \texttt{latticenews}, and sent directly by conference organizers to the conference participants. We have received 92 responses, which is a reasonable response rate for this year's 440 conference participants. We use the R-package \texttt{survey} for evaluation, assuming independent random sampling.
Due to small sample sizes in many underrepresented groups, we need to regroup people of certain characteristics. In particular, we create variables describing a single person as ``diverse''. We choose this word because of its neutrality and mean people who are historically underrepresented or face systematic discrimination. We acknowledge that this choice is not unique and intend to publish further studies examining more effects in the data.
In detail, this regrouping is done as in the following:

\begin{itemize}
\item
  We are introducing the category ``ethnically diverse'' for the answers given for ``Ethnicity'', Here we mapped the ``Category/White'' to ``False'', the answers ``Prefer not to answer'', no answer and answers in the ``others'' that indicated that the respondent did not want to make a statement to the value ``NA''. All other answers were mapped to ``True''.
\item
  For the answers given for gender identity, we mapped again ``Prefer not to answer'', no answer, and answers in ``other'', indicating that the respondent did not want to answer to ``NA''. The answer ``Male'' was mapped to ``False'', and everybody else was mapped to ``True''.
\item
  For the category LGBTQIA+, there was no ``other'' field, so we introduced a boolean that we counted as ``True'' if the response was ``Yes'', as ``False'' if the response was ``No'', and as ``NA'' for ``Prefer not to answer'' and no answer.
\item
  We introduced a category of religious diversity. We mapped a response to ``False'' if the response indicated either no affiliation with any church or Christianity. This choice is, of course, debatable and can be done differently. We plan to additionally study feelings of welcomeness for members of different religious denominations in the future in more detail. We considered slight modifications of this in the ``other field''. If people indicated in the ``other'' field that they did not want to answer, ticked ``Prefer not to answer'' or did not respond at all, we categorized their answer as ``NA''. All other religious denominations were counted as ``diverse''.
\item
  We did not receive enough answers to meaningfully differentiate between visible and invisible disabilities. Due to this we mapped both the answers ``Yes, it is visible to others'' and ``Yes, it is usually not visible to others'' to ``True'', the answer ``Prefer not to answer'' and no answer at all to ``NA'' and everyone who answered ``No'' to ``False''
\item
  In the survey we asked for finer age groups in order to understand the representativeness of the responses for our community. However in the evaluation we only examined generational divides. For this we introduced a category ``Over 40 years old?'' in which we mapped the age groups ``41-50'', ``51-60'' and ``\textgreater{} 60'' to ``True'' and the age groups ``\textless{} 21'', ``21-30'', ``31-40'' to ``False''. No answer and ``Prefer not to answer'' were assigned the value ``NA''.
\item
  Our denominator data is loosely this year's conference participants. While we expect a substantial overlap, there will also be several responses coming from people who participated in the survey and have participated in the conference before but are not participating this year. This also means that we overestimate our response rate and underestimate finite-population effects.
\end{itemize}

\hypertarget{limitations}{%
\subsection{Limitations}\label{limitations}}

We are assuming random sampling from the community. It is probable that there are biases in the sampling since the invitation to do the study was sent out freely over multiple e-mail lists. This means that people who participated had to be subscribed to the list and be interested in participating in the study; people who are interested might respond differently than people who are not. The statistics on ages and diversity representation, however, suggest that our sample is in fact quite close to random sampling within conference attendees.

Nothing prohibited people from completing the survey multiple times, although we have no indication in the data that this has happened.

A random sampling from conference participants however also implies that senior researchers from Europe and North America are overproportionally represented in relation to the full community. This means it is likely that effects for other geographic regions and early-stage researchers are not captured well. It is expected that this means, that diversity-related disparities are underestimated in this study, since it is likely that many scientists who faced problems simply left the field.

\begin{figure}

{\centering \includegraphics{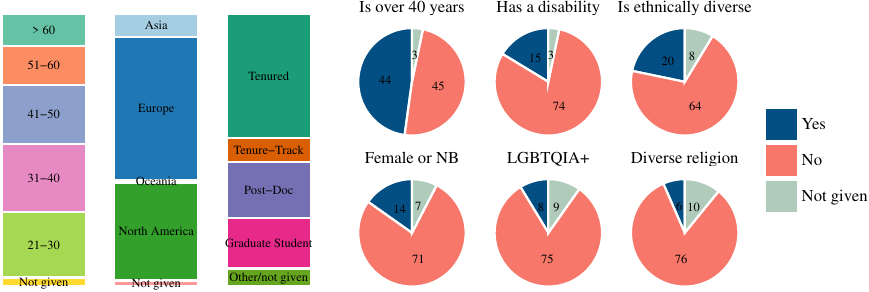} 

}

\caption{Representation of demographic groups in responses}\label{fig:unnamed-chunk-4}
\end{figure}

\hypertarget{feeling-welcome-in-the-community}{%
\section{Feeling Welcome in the Community}\label{feeling-welcome-in-the-community}}

\hypertarget{being-welcome-at-the-conference}{%
\subsection{Being welcome at the conference}\label{being-welcome-at-the-conference}}

People generally report feeling very welcome at the conference, with most people choosing the highest option. However, we found that diverse people feel less welcome than their peers at the core element of the conference: the scientific program and talks. Specifically, ethnically diverse people feel just as welcome at social events and breaks as everybody else but less welcome at talks and the accommodations. People with a diverse religious denomination feel substantially less welcome at all parts of the conference, but the number of responses in this group was overall low.

People with disabilities report on average a high degree of welcomeness, even higher than people without disabilities. There is no significant age divide in feeling welcome to the community, however, people 40 years or younger feel slightly less welcome at social events.

\begin{figure}

{\centering \includegraphics{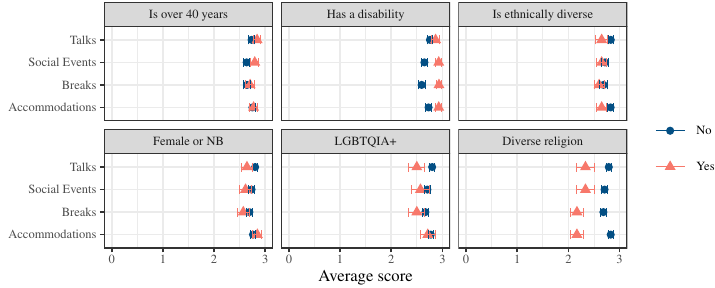} 

}

\caption{Average score to the questions 'Did you feel welcome during/at the... ?'}\label{fig:unnamed-chunk-5}
\end{figure}

\hypertarget{negative-experiences}{%
\subsection{Negative Experiences}\label{negative-experiences}}

We gauged the frequency of negative incidents and experiences by asking respondents to agree or disagree with eleven statements. A neutral answer was possible. For each respondent we then calculated a total, where agreeing to a statement about a negative incident was counted as -1 and agreeing to one about a positive incident was counted as +1. Examining this sum allows us to make a stronger statement than looking at individual statements that might in isolation also leave room for interpretation.

In detail, the respondents were asked to agree or disagree with the following statements:

\begin{itemize}
\tightlist
\item
  During the conference I felt isolated. (-1)
\item
  I had to leave the conference early because of my experience there. (-1)
\item
  During the conference I was actively included in scientific discussion. (+1)
\item
  Sometimes I felt that people were talking down to me. (-1)
\item
  I think that all participants of the conference are afforded the same level of professional respect. (+1)
\item
  Other participants were telling uncomfortable jokes. (-1)
\item
  There was a relaxed atmosphere during the coffee breaks. (+1)
\item
  Someone made a discriminatory comment about me. (-1)
\item
  I was worried about my safety either at the accommodation, or when traveling from or back to the conference site. (-1)
\item
  I had to leave a session early because of my experience there. (-1)
\item
  I avoided conference events because of the people that I knew would be attending. (-1)
\end{itemize}

It was possible to answer ``I don't know'' or not answer the question at all. The value at the end of the statement indicates, whether agreeing to this statement adds or subtracts a point.

\hypertarget{overall-differences}{%
\subsubsection{Overall differences}\label{overall-differences}}

Negative experiences during conference participation are strongly correlated with gender, ethnic background, sexual orientation, religion and age, but we observed no effect with regards to disabilities. The effects observed are qualitatively different. This means that the differences in the scores are not explainable by differences in a single statement but different groups have cited different kinds of negative experiences.

\begin{figure}

{\centering \includegraphics{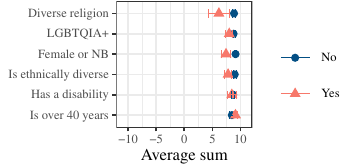} 

}

\caption{Summed scores of negative experiences. The more negative incidents were experienced, the lower the score. Average values are calculated for the sum and compared for different diverse characteristics.}\label{fig:unnamed-chunk-6}
\end{figure}

\begin{figure}

{\centering \includegraphics{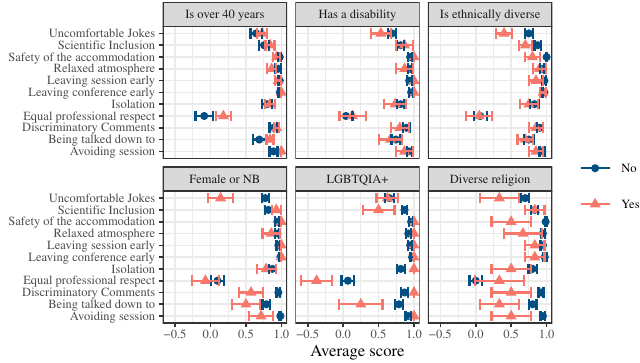} 

}

\caption{Anwers profiles to negative experience statement section. Values are mapped as stated later, higher scores correspond to a positive experience.}\label{fig:answer-profiles}
\end{figure}

\hypertarget{how-people-responded-to-the-statements}{%
\subsubsection{How people responded to the statements}\label{how-people-responded-to-the-statements}}

While strongest evidence for existing inequalities in discussion culture can be seen in the summed scale, we can also examine the origin of these differences in the answers by statement, see figure \ref{fig:answer-profiles}. Note here that these answers provide the additionally interesting insight that for many statements, people belonging to different groups actually do give the same answer on average, but then diverge on very specific aspects. For example, comparing the group of participants below and above 40 years old, we can see, that both groups respond to most categories in the same way on average, however, there are significant differences in their answers to the question whether they think that all participants are awarded the same professional respect, whether they have ever experienced being talked down to or whether they have ever avoided a session because of the people that they knew would be there. Due to the relatively large sample sizes in both groups, we can see that the questionnaire is actually able to express these differences.

Women and non-binary people experience other people making uncomfortable jokes much more frequently than their peers. In addition, they are reporting people making discriminatory comments, being talked down to, and (maybe correspondingly) avoiding sessions because of the people that they knew would be there.

For ethnically diverse people, the effects are different. Here respondents showed the highest discrepancy to their colleagues in feeling included in the scientific discussion and notably feeling safe at the accommodation. They also experience microagressions, such as uncomfortable jokes, but we do not see any differences in their estimation of professional respect, discriminatory comments or being talked down to.

People with an LGBTQIA+ identity report issues with feeling included, receiving the same level of professional respect and being talked down to, but do not report uncomfortable jokes to a higher extent than other respondents.

The data for people with diverse religions is not sufficient to draw conclusions from the individual statements.

\hypertarget{qualitative-answers-in-textboxes}{%
\subsubsection{Qualitative Answers in Textboxes}\label{qualitative-answers-in-textboxes}}

Respondents further elaborated negative experiences they had at the conference in comments. Here people mostly described a hostile environment, mentioning ``belittling jokes'', instances of being ``talked down'' to, ``ad hominem remarks'', ``fights'' and people ``generally not being polite''. One PhD student described that a senior researcher laughed at the answer they gave him and a researcher described that their results were not believed, questions were adressed to their supervisor instead of directly to them, and they were socially excluded at the conference.
Concrete microagressions were also described, for example making disparaging remarks about the country of origin of participants. One respondent describes that they feel that Asians are being regarded as second class citizens in Europe.

We have to conclude that even from this limited data, we can see problems in the discussion culture at the conference. It is vital that this is improved by continuing to employ the code of conduct and develop a comprehensive response to breaches. The latter is especially of high importance since people describe microaggressions and racism at the conference.

Organizing mentoring and networking events as part of the conference can facilitate exchange between participants of different nationialities and age ranges, thereby helping to remove the barriers between them.

\hypertarget{becoming-informed}{%
\section{Becoming informed}\label{becoming-informed}}

\hypertarget{estimating-data}{%
\subsection{Estimating Data}\label{estimating-data}}

The community is struggling with acknowledging inequities. We can see in the data that female representation at the conference is consistently overestimated by an approximate factor of 2, a similar value that was found in the 2019 survey, see \citep{Aubin:2019rdf}.

\begin{figure}

{\centering \includegraphics{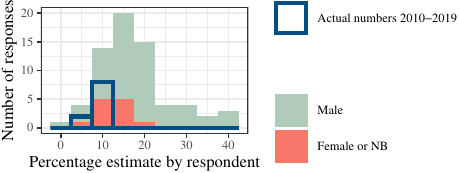} 

}

\caption{Perception of female representation at the conference depending on the gender of the respondent}\label{fig:unnamed-chunk-7}
\end{figure}

This divide in estimating the problem correctly can also be found in the observation of discrimination, harassment and exclusion of people due to their gender identity, expression or sexual orientation. Here either observing or experiencing incidents was reported more often by diverse respondents. It seems, that women and non-binary people are seeing incidents that their male peers are apparently overlooking.

\hypertarget{effectivity-of-sources}{%
\subsection{Effectivity of sources}\label{effectivity-of-sources}}

For the information that people consult to inform themselves about inequity in the community, gender diverse people are consulting a broader amount of sources. It is encouraging to see that there does not seem to be an age divide between the consulted sources. The pattern of sources consulted is the same across ages.

\begin{figure}

{\centering \includegraphics{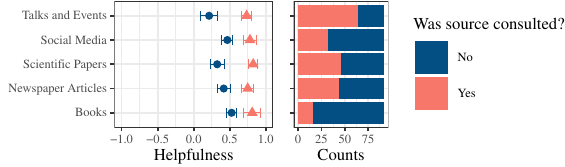} 

}

\caption{Counts of respondents having consulted a certain sources and how helpful they perceived it, from -1 very unhelpful to +1 very helpful.}\label{fig:unnamed-chunk-10}
\end{figure}

People feel most likely enabled to take action against inequity after reading either scientific papers, listening to talks or participating in events. Talks and events generally seem to be a great source of information for the community; most of the community have attended one at some point. Overall, as one would expect, people who marked many different sources of information are more likely to say that it was helpful, than those who consulted less.

\hypertarget{recommendations-and-follow-up}{%
\section{Recommendations and Follow-up}\label{recommendations-and-follow-up}}

We follow up on the following recommendations from our survey published in \citep{Aubin:2019rdf}. The following tasks were suggested:

\begin{enumerate}
\def\labelenumi{\arabic{enumi}.}
\tightlist
\item
  \emph{Provide a code of conduct and diversity surveys organized around the conference.}
\end{enumerate}

A code of conduct has been developed and employed at the conference. It is being reviewed and adapted regularly.

This is the third survey since 2019 on diversity and inclusivity. We aim to conduct surveys every year with different focuses.

\begin{enumerate}
\def\labelenumi{\arabic{enumi}.}
\setcounter{enumi}{1}
\tightlist
\item
  \emph{Plenary at the conference on diversity and inclusivity and actively increase diversity of physics plenary speakers and session chairs}
\end{enumerate}

Lattice 2023 had multiple speakers on diversity. Diverse scientists held plenary talks and chaired the scientific program.

\begin{enumerate}
\def\labelenumi{\arabic{enumi}.}
\setcounter{enumi}{2}
\tightlist
\item
  \emph{Financial support for minorities and students}
\end{enumerate}

Financial support is so far provided in terms of conference fee reductions or waivers. Improvements to reduce barriers further, especially for (early-stage) researchers with low funding opportunities, are in progress --- additionally, the lattice 2024 LOC plans to provide and organize support for on-site child care.

\begin{enumerate}
\def\labelenumi{\arabic{enumi}.}
\setcounter{enumi}{3}
\tightlist
\item
  \emph{Mentoring program at the conference to improve networking of diverse early stage researchers}
\end{enumerate}

The women's lunch has been changed into a diversity lunch, providing a networking opportunity for all diverse participants at the conference.

\begin{enumerate}
\def\labelenumi{\arabic{enumi}.}
\setcounter{enumi}{4}
\item
  \emph{Promote international exchange between different career stages}
\item
  \emph{Development of a codified and transparent reporting mechanism for breaches of the code of conduct}
\end{enumerate}

This is a work in progress by the current committee.

\begin{enumerate}
\def\labelenumi{\arabic{enumi}.}
\setcounter{enumi}{6}
\tightlist
\item
  \emph{Provide a practical diversity and inclusivity guide for the conference organizers to follow to promote a positive atmosphere}
\end{enumerate}

\hypertarget{conclusion}{%
\section{Conclusion}\label{conclusion}}

We found differences in feelings of welcomeness between respondents belonging to a diverse group and their peers, where being a woman or non-binary, identifying as LGBTQIA+, members of diverse religious denominations, and ethnically diverse people felt less welcome than their colleagues. Overall, these differences are small but, given a few assumptions, statistically significant. While the sample size was limited and we expect biases to be present in the data, there are qualitative effects visible that are useful for developing recommendations to improve the atmosphere at the conference and make it more inclusive. This is crucial since our specialization struggles with diversity, even compared to other specializations within physics.

\hypertarget{acknowledgements}{%
\section{Acknowledgements}\label{acknowledgements}}

We thank Andrea Bossmann from the FU Berlin for useful comments that improved our questionnaire. WD is supported by U.S.~Department of Energy, Office of Science, Office of Nuclear Physics under contract number DE-SC0011090 and DE-SC0023116 and by the US National Science Foundation under Cooperative Agreement PHY-2019786. SM is funded by the European Union’s Horizon 2020 research and innovation program under the Marie Skłodowska-Curie grant agreement \textnumero~813942. We thank Martha Constantinou, Luigi Del Debbio, Liuming Liu, Antonin Portelli, Indrakshi Raychowdhury and Fernando Romero-López for consultation during this study.

\renewcommand\refname{References}
  \bibliography{literature.bib}

\begin{thebibliography}{}

\bibitem[Aubin et~al., 2019]{Aubin:2019rdf}
Aubin, C., Bali, G., Del~Debbio, L., Detmold, W., G\"ulpers, V., Hollitt, S.,
  Lin, H.-W., Liu, L., and Ryan, S.~M. (2019).
\newblock {Report on the 2019 Lattice Diversity and Inclusivity Survey}.
\newblock {\em PoS}, LATTICE2019:295.

\bibitem[Aubin et~al., 2021]{poster:2021}
Aubin, C., Bali, G., Del~Debbio, L., Detmold, W., G\"ulpers, V., Hollitt, S.,
  Lin, H.-W., Liu, L., and Ryan, S.~M. (2021).
\newblock Report on the 2021 lattice diversity and inclusion survey.

\bibitem[Lin, 2017]{Lin:2016age}
Lin, H.-W. (2017).
\newblock {Some Statistics on Women in Lattice QCD}.
\newblock {\em PoS}, LATTICE2016:366.

\end{thebibliography}

\end{document}